\documentstyle[prb,aps,epsf]{revtex}
\draft

\begin{document}

\title{Conductance anomalies in a one-dimensional quantum contact}

\author{O.P. Sushkov}

\address{School of Physics, University of New South Wales,\\
 Sydney 2052, Australia}

\maketitle

\begin{abstract}
Short length quantum wires (quantum contacts) exhibit a conductance structure
at the value of conductance close to $0.7 \times 2e^2/h$. 
The structure is also called the conductance anomaly.
In longer contacts the  structure evolves to the lower values of conductance.
We demonstrate that this structure is related to the development of charge
density waves within the contact. This is a precursor for Wigner 
crystallization. Many-body Hartree-Fock calculations of conductance are
performed. The results are in agreement with experimental data.
\end{abstract}


\section{Introduction}
The quantized conductance $G=nG_2$, $n=1,2,3,...$, $G_2=2e^2/h$,
through a narrow quantum point contact was
discovered in 1988 \cite{Wharam,Wees}. This quantization can be  
understood within a one-dimensional (1D)
non-interacting electron gas picture, see e.g. Ref.\cite{But}.
In the present work we are interested in a deviation from the
integer quantization. This deviation, the  so called 
``0.7 anomaly'' has been found in experimental works \cite{Thomas,Thomas1}.
The anomaly is a shoulder-like feature or a narrow plateau at
$G\approx 0.7G_2$. The recent work \cite{Reilly} demonstrates
that the anomaly evolves down to $G\approx 0.5G_2$
in longer quantum contacts.
Another highly interesting recent experimental
development  is an indication  of the above 
barrier excitation \cite{Kristensen}. The excitation is probably related to 
the ``0.7 anomaly''.

There is practically no doubt that the structure is due to the
interaction between the electrons.
There have been numerous attempts to explain the ``0.7 anomaly'' by spontaneous
magnetization of the 1D quantum wire due to exchange Coulomb interaction
\cite{Chuan,Calmels,Zabala,V'yurkov}.
A somewhat similar, but still different idea is to explain  
the anomaly by formation of a two-electron bound state with total spin $S=1$,
see Refs. \cite{Flamb,Rejec}.

It is not clear why the spontaneous magnetization scenario  can be valid
since it contradicts to the rigorous Lieb-Mattis theorem \cite{Lieb}.
The theorem claims that any 1D many-body system with quadratic kinetic energy 
and with a potential interaction has the ground state with zero spin. 
The same argument is applicable to  the S=1 bound state picture. 
Strictly speaking in the later case one can avoid a formal contradiction with
the theorem saying that the bound state is localized somewhere in
the transition region between the 2D or 3D conductor and the 1D wire.
However this scenario seems unlikely and in the present work I consider
only S=0 ground state in accordance with Lieb-Mattis theorem.

The effect of Wigner crystallization in low density 2D and 3D electron gas
 is well known. There is no such a thing as an ideal  1D Wigner crystal 
because quantum fluctuations destroy any long range order. However it was 
pointed out some time ago \cite{Glaz} that impurities suppress the 
fluctuations and pin the 1D Wigner crystal. Surprisingly this happens at 
pretty high density of electrons $n$: $n \approx 0.5/a_B$,
where $a_B$ is the Bohr radius.
A quantum contact (=short quantum wire) is not uniform and hence it pins
the crystal. This must result in nontrivial behavior of the contact 
conductance.
It is the effect we study in the present work using  the Hartree-Fock 
(HF) method. We demonstrate that the effect results in the conductance anomaly 
at $G \approx 0.7G_2$ for a short contact and in several anomalies for 
longer contacts. In the present work we consider only zero temperature
case. However in conclusions we comment on temperature dependence.

The idea of relation between the conductance anomaly and the Wigner
crystallization was also suggested in a recent paper \cite{Spivak}.
Authors of this paper discuss a ferromagnetic state embedded in
the Wigner crystal. 
We follow the different way considering the ground state with total spin zero
in accordance with Lieb-Mattis theorem.
 
In independent particle approximation, i.e. in the case of an ideal electron
gas, a calculation of the conductance for a given transverse channel is 
straightforward. Assuming adiabatic transmission one gets  
\begin{equation}
\label{GT}
G={{2e^2}\over{h}} T,
\end{equation}
where $T$ is the barrier transmission probability at Fermi energy 
\cite{Land,But}.
In the case of interacting particles this formula is also valid because
before and after the potential barrier the density of electrons
is high enough, and hence the interaction is negligible.
However one can not use the single particle description to 
calculate the transmission probability $T$ because in the vicinity of the 
barrier the electron density is low, and hence the many-body effects are very 
important.
The main question is how to calculate the transmission probability $T$?
To do this we apply the following trick. Consider the liquid of
electrons on a 1D ring with a potential barrier somewhere on the ring.
There is no a current in the ground state of the system. Now let us apply a
magnetic flux through the ring. This flux induces the electric current.
On the one hand the current is related to the barrier transmission 
probability $T$. On the other hand, the current can be calculated using 
the HF method. This allows us to find $T$ with account of many-body effects.

Structure of the present paper  is the following: 
The relation between the current on the ring and the transmission probability
is derived in Sec. II.
In Sec. III we demonstrate how the Hartree-Fock method describes development
of the charge-density wave at low electron density.
The results of selfconsistent calculations of the transmission probabilities
for different barriers are presented in Sec. IV.
Sec. V summarizes the results and presents our conclusions.

\section{Relation between the barrier transmission probability and the
 current in the ring induced by the magnetic flux.}
Let us consider a single particle in a potential $U(x)$. The potential
is localized near $x=0$,  $U(x) =0$ at 
$|x|>l/2$. For simplicity we assume that the potential is symmetric, 
$U(x)=U(-x)$. In the scattering problem the wave function outside the region
of the potential is of the form
\begin{eqnarray}
\label{psi1}
&&x < -l/2: \ \ \ \chi = e^{ikx}+Ae^{-ikx},\\
&&x > l/2
: \ \ \ \chi = B e^{ikx},\nonumber
\end{eqnarray}
where $k=\sqrt{2E}$ is the momentum of the particle, $A$ is the reflection
amplitude and $B$ is the transmission amplitude. Throughout the paper
we set the electron mass, electric charge, and the Planck constant equal to 
unity, $m=e=\hbar =1$. The transmission probability is $T=|B|^2$.
One can also use the basis of standing waves. Since the potential is
symmetric, there is one symmetric and one antisymmetric
solution at a given energy $E$
\begin{eqnarray}
\label{psi2}
&&x < -l/2: \ \ \ \chi_{\pm} = \cos(kx+\delta_{\pm}),\\
&&x > l/2: \ \ \ \chi_{\pm} = \pm \cos(kx - \delta_{\pm}).\nonumber
\end{eqnarray}
Comparing eqs. (\ref{psi1}) and (\ref{psi2}) one can express the 
transmission and reflection amplitudes in terms of the scattering phases
\begin{eqnarray}
\label{ABT}
&&A=e^{-i\Delta}\cos\delta,\nonumber\\
&&B=-ie^{-i\Delta}\sin\delta,\\
&&T=|B|^2=\sin^2\delta,\nonumber
\end{eqnarray}
where
\begin{eqnarray}
\label{Dd}
&&\Delta=\delta_++\delta_-,\\
&&\delta=\delta_+-\delta_-.\nonumber
\end{eqnarray}

Now we switch the vector potential ${\cal A}$ on
\begin{equation}
\label{H1}
H={{(p-{\cal A})^2}\over{2}}+U(x),
\end{equation}
and close the ring imposing the periodic boundary condition
\begin{equation}
\label{per}
\psi(-L/2)=\psi(L/2).
\end{equation}
 We will assume that the length of the
ring $L$ is much larger than the size of the region where the
potential $U(x)$ is nonzero: $L \gg l$. Note that the vector 
potential we introduce is the pure gauge one. We need it to
induce the orbital current in the ring. The magnetic field is zero.
The induced current depends only on the Bohm-Aharonov phase
$\phi={\cal A}L$. In the many-body problem the current has period
$\Delta\phi=\pi$. The current is maximum at 
$\phi=\pm\pi/2,\pm3\pi/2,...$. For further considerations we take
the first maximum
\begin{equation}
\label{AA}
{\cal A}=\pi/2L, \ \ \ \phi=\pi/2.
\end{equation}

Solution of the Schr\"{o}dinger equation with Hamiltonian (\ref{H1}),(\ref{AA})
and the periodic boundary condition (\ref{per}) gives  the following
single electron wave functions outside of the region of the potential.
\begin{equation}
\label{psi3}
\psi_{\pm}(x)={1\over{\sqrt{L(1+|a_{\pm}|^2)}}}
\left(e^{-i\Delta}e^{ikx}+a_{\pm}e^{-ikx+2i{\cal A}x}\right)
\end{equation}
Here $l/2 < x < L-l/2$,
\begin{equation}
\label{apm}
a_{\pm}={{1\mp ie^{-i\delta}}\over
{1\pm ie^{-i\delta}}},
\end{equation}
and the momentum $k$ obeys the following quantization condition
\begin{equation}
\label{quant}
e^{ikL}=\pm ie^{i\Delta}.
\end{equation}
There are two solutions: $\psi_+$ and $\psi_-$.
To elucidate meaning of these solutions let us look at two limiting cases.
The first limit is an infinitely high potential barrier, $U\to \infty$.
In this case the scattering phases 
are equal, $\delta_+=\delta_-$. Substituting these phases into eqs. 
(\ref{apm}), (\ref{psi3}) one finds that the wave functions $\psi_{\pm}$ 
coincide with symmetric and antisymmetric standing waves (\ref{psi2}).
In the second limit there is no barrier, $U=0$. Then the scattering phases 
are the following: $\delta_+=0$, $\delta_-=-\pi/2$. One can prove that in 
this case the states  (\ref{psi3}) describe clockwise and anti-clockwise 
rotations,  $\psi_+\to \exp(ikx)$, $\psi_-\to \exp(-ikx+2iAx)$.

The quantization condition (\ref{quant}) gives the following momenta
\begin{equation}
\label{kk}
k_m^{(\pm)}={{2\pi}\over{L}}(m\pm1/4+\Delta/2\pi),\ \ \ \ 
 m=0,1,2,3...,
\end{equation}
and energy levels
\begin{eqnarray}
\label{ee}
&&\epsilon_m^{(+)}={1\over{2}}\left({{2\pi}\over{L}}\right)^2
\left(m+\Delta/2\pi\right)^2, \\
&&\epsilon_m^{(-)}={1\over{2}}\left({{2\pi}\over{L}}\right)^2
\left(m-1/2+\Delta/2\pi\right)^2.\nonumber
\end{eqnarray}
Using the wave functions (\ref{psi3}) one easily finds the electric
current in each single particle state
\begin{eqnarray}
\label{jj}
&&J_m^{(+)}={{2\pi}\over{L^2}}\left(m+\Delta/2\pi\right)\sin\delta, \\
&&J_m^{(-)}=-{{2\pi}\over{L^2}}\left(m-1/2+\Delta/2\pi\right)\sin\delta.
\nonumber
\end{eqnarray}

Up to now we were considering the single particle problem. Now let us 
consider the many body problem. To do this we fill all the single particle 
states up to the Fermi energy. Each orbital state is filled by two electrons,
spin up and spin down.
Electric currents $J_m^{(+)}$ and $J_m^{(-)}$
partly compensate each other, so the total electric current of the
many-body system is
\begin{eqnarray}
\label{jjj}
J&=&2\sum_m\left(J_m^{(+)}+J_m^{(-)}\right)=
{{4\pi}\over{L^2}}\sum_m\left[(m+\Delta/2\pi)
(\left(\sin\delta(k_m^{(+)})-\sin\delta(k_m^{(-)})\right)+
{1\over{2}}\sin\delta(k_m^{(-)})\right]\nonumber\\
&=&{{2\pi}\over{L^2}}\sum_m\left[k{{d\sin\delta}\over{dk}}+\sin\delta\right]
={{2\pi}\over{L^2}}\sum_m {{d(k\sin\delta)}\over{dk}}
={{1}\over{L}}\int_0^{k_F}{{d(k\sin\delta)}\over{dk}}dk=
{{k_F}\over{L}}\sin\delta_F={{\pi N}\over{2L^2}}\sin\delta_F.
\end{eqnarray}
Here $N$ is the number of electrons on the ring, $k_F=\pi N/2L$ is the
Fermi momentum, and $\delta_F$ is the scattering phase $\delta$, see eq.
(\ref{Dd}), taken at the Fermi energy.

Eq. (\ref{jjj}) gives the main tool we use in the present work.
First we solve numerically  the many body problem on the ring without the 
barrier, $U=0$, and find the electric current $J_0$. Second we solve 
numerically the many body problem with the barrier, $U\ne 0$, and find the 
electric  current $J_U$. Then using eqs. (\ref{jjj}) and (\ref{ABT}) we 
conclude that the transmission probability at the Fermi energy is
\begin{equation}
\label{TT}
T=\left(J_U/J_0\right)^2.
\end{equation}
This is exactly the transmission probability we need for the conductance
(\ref{GT}).

In the derivation of eqs. (\ref{jjj}) and (\ref{TT}) we neglect 
electron-electron interaction outside the potential barrier. This is 
valid if the density of electrons is high enough.
In the vicinity of the barrier the electron density is low and hence
the electron-electron interaction is very important. This does not
contradict to the presented derivation. The information about the 
electron-electron interaction is hidden in the scattering phases.

Finally we would like to comment on the Luttinger liquid behavior
that is related to the long-range fluctuations in the 1D interacting
Fermi gas. The HF method we use to analyze the electron
ring certainly does not take into account the long-range fluctuations.
However this is irrelevant to the calculation of the
barrier transmission probability.
The matter is that the number of electrons above the quantum barrier
is not more than several, so there is no room for the long-range
fluctuations within the barrier. The long ring outside the barrier
 is just a technical trick to describe the reservoir of 
electrons. There is no need to take into account the long-range fluctuations
in the reservoir.

\section{Hartree-Fock (HF) method  for interacting 
electrons on the ring in the external gauge field.
 Charge-density wave pinned by the impurity.}
The Hamiltonian of the many body system we consider is of the form
\begin{equation}
\label{Hmb}
H=\sum_i\left[{{(p_i-{\cal A})^2}\over{2}}+U(x_i)\right]+{1\over{2}}\sum_{i,j}
V(x_i,x_j),
\end{equation}
where indexes $i$ and $j$ numerates electrons, $x_i$ is the periodic
coordinate ($0<x<L$), ${\cal A}$ is the gauge field (\ref{AA}), 
$U(x)$ is the 
potential barrier, and $V_{ij}$ is the electron-electron interaction.
We use atomic units, so coordinates are measured in unites of Bohr
radius, $a_B=\epsilon\hbar^2/m e^2$, and energies are measured in units
of $E_{unit}=me^4/\hbar^2\epsilon^2$, where $m$ is the effective electron
mass and $\epsilon$ is the dielectric constant. 
For experimental conditions of works \cite{Thomas,Thomas1,Reilly,Kristensen} 
these values
are the following: $a_B \approx 10^{-2}\mu m$, $E_{unit}\approx 10^{-2}eV$.
The electron-electron Coulomb repulsion we take in the form
\begin{equation}
\label{Vij}
V(x,y)={1\over{\sqrt{a_t^2+D^2(x,y)}}},
\end{equation}
where $a_t$ is the effective width of the transverse channel, and $D(x,y)$ is
the length of the shortest arc  between the points $x$ and $y$
on the ring, see comment \cite{DL}. 
The transverse motion of electrons is frozen due to some confining
potential that can be approximated as
\begin{equation}
\label{conf}
U_{\perp}(\rho)={{m\omega_{\perp}^2\rho^2}\over{2}}.
\end{equation}
According to the data \cite{Kristensen,Reilly} the energy splitting
between the transverse channels is $\hbar \omega_{\perp} \sim 4-7meV$.
This gives the following value of the first channel transverse size 
\begin{equation}
\label{at}
a_t = 1/\sqrt{m\omega_{\perp}} \approx 2a_B.
\end{equation}
Averaging over the oscillator wave function of the transverse motion
shows that the effective 1D Coulomb interaction is well approximated
by eq. (\ref{Vij}).
We stress that the value of $a_t$ is close to the critical distance
between the electrons at which the Wigner crystallization occurs
\cite{Glaz}. This is a fortunate coincidence. Larger value of $a_t$ would 
suppress the conductance anomaly, and this really happens in higher transverse
channels. Substantially smaller value of $a_t$ would require a theoretical 
technique more sophisticated than the HF
method employed in the present work. We will discuss this point later.
Let us also comment on the long-range behavior of the interaction (\ref{Vij}).
Because of the ring geometry the interaction (\ref{Vij}) at $|x-y|\sim L$
is somewhat different from the Coulomb one. However the so long range tail
of the interaction does not influence the results. Anyway in a real system
there is a screening by gates and many other effects that influence the long
range tail. What is important is that at the size of the barrier
$|x-y|\sim l \ll L$ the interaction (\ref{Vij}) has the right form,
$V=1/|x-y|$.

In the HF method the many body wave function of the system 
(\ref{Hmb}) is represented in the form of Slater determinant of single 
particle orbitals  $\varphi_i(x)$. Each orbital obeys the equation
\begin{equation}
\label{hf}
\hat{h}\varphi_i=\epsilon_i\varphi_i,
\end{equation}
where $\epsilon_i$ is the single particle energy and
$\hat{h}$ is the HF Hamiltonian
\begin{eqnarray}
\label{Hhf}
\hat{h}\varphi(x)&=&\left({{(p-{\cal A})^2}\over{2}}+U_{eff}(x)
\right)\varphi(x)
-\sum_j\int\varphi_j^*(y)\varphi(y)V(x,y)dy\varphi_j(x),\\
U_{eff}&=&U(x)+ \sum_j\int|\varphi_j(y)|^2 V(x,y)dy.\nonumber
\end{eqnarray}
Here the summation is performed over all filled orbitals.
To avoid misunderstanding we have to comment on a terminological 
question. Usually only the case of zero vector potential, ${\cal A}=0$,
is called the HF method. If ${\cal A}\ne 0$ the method is
called the Hartree-Fock method in the external field, or
 if ${\cal A}$ is time dependent it is called the time dependent 
Hartree-Fock 
method \cite{TDHF}. The external field method is equivalent to
the Random Phase Approximation (RPA).
A crucial point is that the external field Hartree-Fock method (=RPA)
is gauge invariant,  and therefore the electric current calculated
within this method is conserved  \cite{Dzuba}.

For computations we use a ring of length $L=80$ with a finite grid of
400 or 600 points. A naive finite grid version of the Hamiltonian 
(\ref{Hhf}) has a weak dependence on the gauge and hence the
electric current is not exactly conserved. To avoid this minor
trouble we use the lattice (=grid) modification of the 
Hamiltonian (\ref{Hhf}) replacing $(p-{\cal A})^2\varphi$ by 
$\left[2|\varphi(n)|^2-\varphi^*(n+1)e^{iAh}\varphi(n)
-\varphi^*(n)e^{-iAh}\varphi(n+1)\right]/2h^2$. Here $h$ is the step of the
grid and $\varphi(n)$ is the wave function on the site $n$ of the grid.
The electric current corresponding to the grid Hamiltonian is
\begin{equation}
\label{jg}
J=-\sum_j{{i}\over{2h}}\left[\varphi_j^*(n)e^{iAh}\varphi_j(n+1)-
\varphi_j^*(n+1)e^{-iAh}\varphi_j(n)\right].
\end{equation}
This current is exactly conserved at a finite $h$, and at $h\to 0$ it
coincides with the standard electric current.

First we consider the impurity pinned Wigner crystallization \cite{Glaz}.
The electron density on the ring found as a result
of the self consistent solution of the HF equations 
without any impurity, $U(x)=0$, is plotted in Fig.1 by long dashed lines.
We consider the total number of electrons 
$N=N_{\uparrow}+N_{\downarrow}=78, 50, 38, 18$,
$N_{\uparrow}=N_{\downarrow}$.
The density is homogeneous. The density with the external
potential
\begin{equation}
\label{UU}
U(x)={{U_0}\over{e^{(|x|-l/2)/d}+1}}
\end{equation}
at $U_0=0.5$, $l=8$, and $d=1$ are shown by solid lines.     

\begin{figure}[h]
\vspace{-10pt}
\hspace{-35pt}
\epsfxsize=11cm
\centering\leavevmode\epsfbox{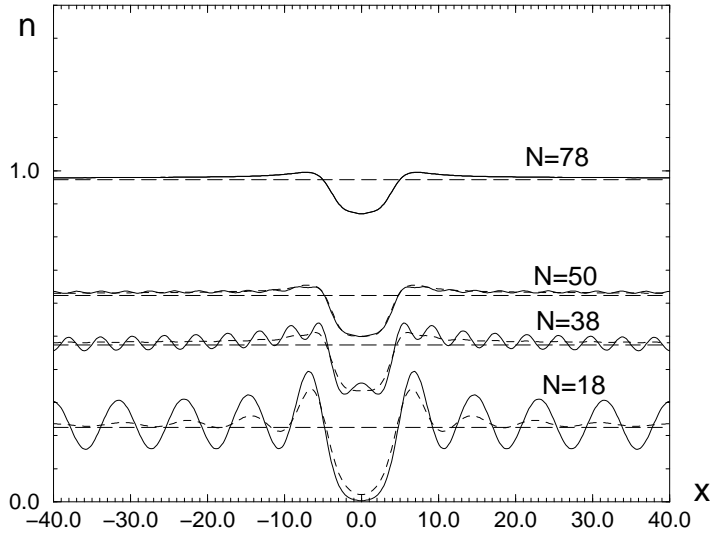}
\vspace{-10pt}
\caption{\it {The Hartree-Fock electron density on the ring without 
(long dashed line) and with (solid line) small impurity.
The dashed line corresponds to the Hartree approximation.
The number of electrons on the ring is $N=78,50,38,18$.}}
\label{Fig1}
\end{figure}
\noindent
It is clearly seen that the impurity pinned charge density wave
is developed at the electron density $n \approx 0.5$. For comparison, we 
also show by dashed lines the results obtained in the Hartree approximation
(no exchange interaction). The Hartree approximation also gives the
charge density wave  but underestimates the effect. In real Wigner crystal
spins are localized at the sites of the lattice. This spin structure
certainly can not be obtained in the HF approximation that enforces
zero spin density at any point. However we are not going to study
a real Wigner crystal at very low electron density. Anyway such a 
state does not conduct an electric current. We are interested 
in $n\geq 0.5$. The charge density wave at $n\sim 0.5$ is only a precursor 
for the Wigner crystallization and this precursor can be described in the
HF approximation. 

In obtaining the results presented on Fig.1 we have used the transverse
cutoff $a_t=2$, see eqs. (\ref{Vij}),(\ref{at}).
At the smaller value of $a_t$ the charge density wave is developed at
a higher electron density. This reflects the fact that the HF
method with the ideal Coulomb interaction overestimates the tendency
towards the charge density waves. Fortunately the value of $a_t\sim 2$
compensates this overestimation giving the correct value of the
``critical'' density \cite{Glaz}. This is why the application of the 
Hartee-Fock method is justified.

We also performed HF calculations with total spin $S\ne 0$.
Technically it is very simple, we just set $N_{\uparrow}\ne N_{\downarrow}$.
The macroscopic magnetization, i.e. $S\sim N$, does not arise.
However, it is interesting to note that at the low electron density,
$n \leq 1$, the ground state with $S=1$ or $S=2$ sometimes has slightly
lower energy than the state with $S=0$. 
This clearly contradicts to the rigorous Lieb-Mattis theorem \cite{Lieb} 
and therefore it is just a byproduct  of the HF approximation.
We disregard this result and consider only the states with $S=0$.

\section{Many-body calculation of the barrier transmission probability}
For calculations in this section we take the number of electrons on
the ring $N=N_{\uparrow}+N_{\downarrow}=158$,
$N_{\uparrow}=N_{\downarrow}$. This corresponds to the average electron
density $<n>= 1.975$ that is well above the charge density wave threshold.
The external potential barrier is taken in the form (\ref{UU}).
The parameter $U_0$ models the gate potential in the experiments
\cite{Thomas,Thomas1,Reilly,Kristensen}.
Selfconsistent solution of the HF eqs. (\ref{hf}) is performed 
at $U_0=0$ and at $U_0\ne 0$ and then the
transmission probability is found using eq. (\ref{TT}).
The transmission probability as a function of $U_0$ at $d=0.5$ and 
the values of the barrier length
$l=4,6,8,10,12$ is plotted in Fig.2.
\begin{figure}[h]
\vspace{-10pt}
\hspace{-35pt}
\epsfxsize=10.7cm
\centering\leavevmode\epsfbox{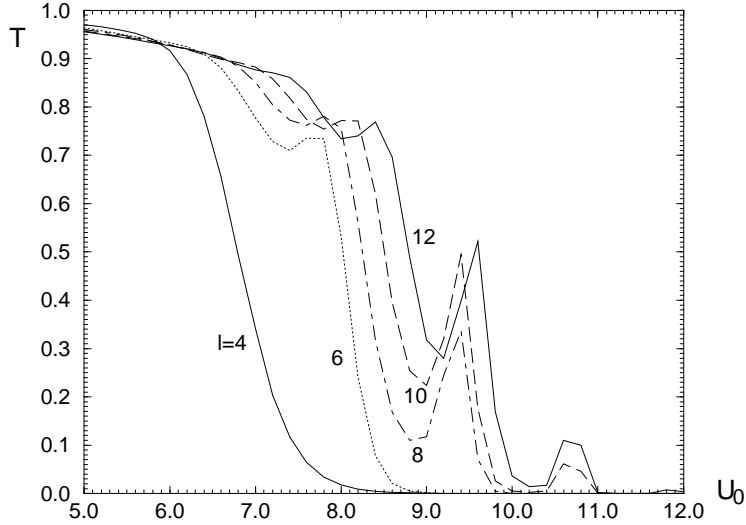}
\vspace{-10pt}
\caption{\it {The transmission probability $T$ versus the
potential $U_0$. The transmission probability has been calculated
in the Hartree-Fock approximation at $d=0.5$ for the values of
the barrier length $l=4,6,8,10,12$.}}
\label{Fig2}
\end{figure}
\noindent
The plot of $T$ versus $U_0$ is representative if $L\gg l$. However
in the computations $L=80$ and therefore the inequality is not so
strong. As a result the Fermi energy attains a weak $U_0$ dependence. 
To compensate this effect, in Fig.3 we plot the transmission
probability $T$ versus the value of 
$V_0=U_0-\epsilon_F(U_0)+\epsilon_F(U_0=0)$
that models the gate potential more accurately.
\begin{figure}[h]
\vspace{-10pt}
\hspace{-35pt}
\epsfxsize=10.7cm
\centering\leavevmode\epsfbox{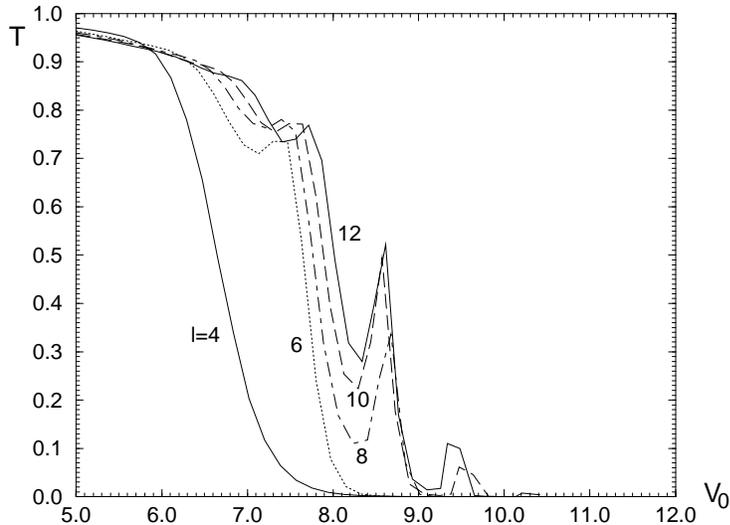}
\vspace{-10pt}
\caption{\it {The transmission probability $T$ versus the gate
potential $V_0$. The transmission probability has been calculated
in the Hartree-Fock approximation at $d=0.5$ for the values of
the barrier length $l=4,6,8,10,12$.}}
\label{Fig3}
\end{figure}
\noindent
There is no a qualitative difference between Fig.2 and Fig.3, but
 Fig.3 is more correct quantitatively.
There are some  very sharp structures in Fig.2 and Fig.3, but they arise only
because of the relatively large step in $U_0$. The computations are
rather time consuming and this limits the number of points.
For comparison we present in Fig.4 the same plots as in Fig.3, but
the curves have been obtained in the Hartree approximation. So the exchange
term in eq. (\ref{Hhf}) has been dropped out.
\begin{figure}[h]
\vspace{-10pt}
\hspace{-35pt}
\epsfxsize=10.7cm
\centering\leavevmode\epsfbox{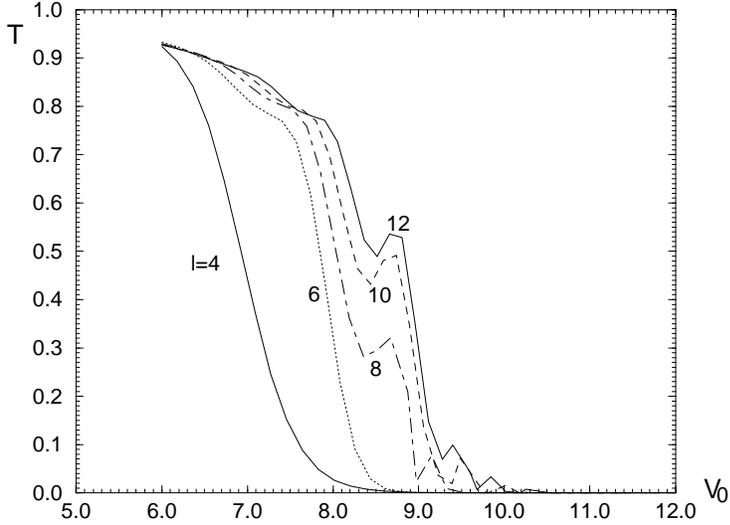}
\vspace{-10pt}
\caption{\it {The transmission probability $T$ versus the gate
potential $V_0$. The transmission probability has been calculated
in the Hartree approximation at $d=0.5$ for the values of
the barrier length $l=4,6,8,10,12$.}}
\label{Fig4}
\end{figure}
\noindent
The results of Hartree approximation are qualitatively
similar to that of the HF approximation, Fig.3. However
all the structures are less pronounced.

The previous figures present the barrier transmission probability
calculated at the parameter $d=0.5$, see eq. (\ref{UU}). 
In Fig.5 and Fig.6 we plot the similar results at $d=1$. The Fig.5 
corresponds to the  HF approximation and Fig.6 corresponds 
to the Hartree approximation.
\begin{figure}[h]
\vspace{-10pt}
\hspace{-35pt}
\epsfxsize=10.7cm
\centering\leavevmode\epsfbox{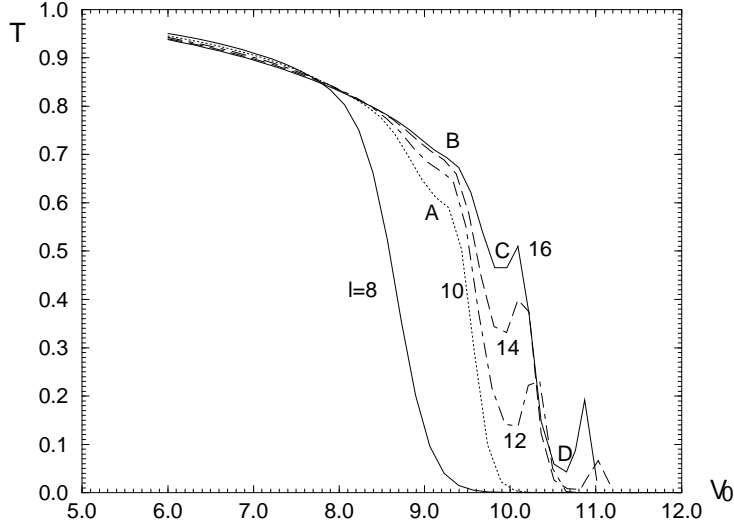}
\vspace{-10pt}
\caption{\it {The transmission probability $T$ versus the gate
potential $V_0$. The transmission probability has been calculated
in the Hartree-Fock  approximation at $d=1$ for the values of
the barrier length $l=8,10,12,14,16$.}}
\label{Fig5}
\end{figure}
\noindent

\begin{figure}[h]
\vspace{-10pt}
\hspace{-35pt}
\epsfxsize=10.7cm
\centering\leavevmode\epsfbox{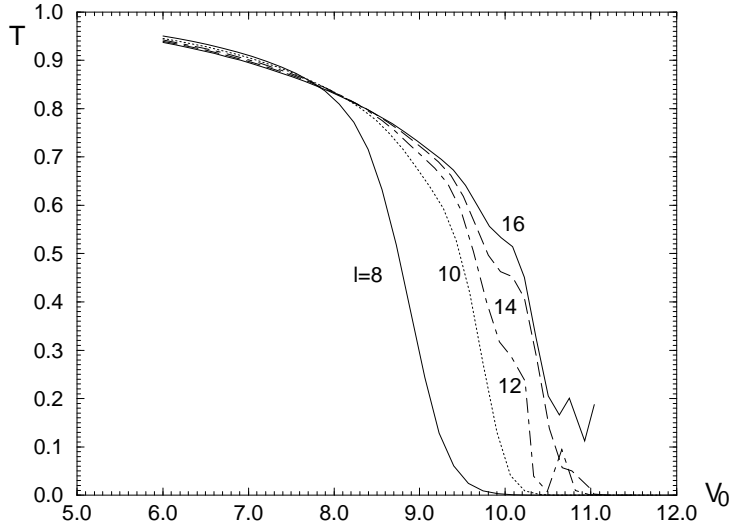}
\vspace{-10pt}
\caption{\it {The transmission probability $T$ versus the gate
potential $V_0$. The transmission probability has been calculated
in the Hartree  approximation at $d=1$ for the values of
the barrier length $l=8,10,12,14,16$.}}
\label{Fig6}
\end{figure}
\noindent
The Hartree approximation gives results qualitatively
similar to that of the HF approximation, but the
Hartree approximation clearly underestimates the structure.

Fig.3 and Fig.5 clearly demonstrate a shoulder at the value of the
transmission probability $T\sim 0.7$. For longer barriers an additional 
structure appears at smaller values of $T$. This is in agreement with
experimental data \cite{Thomas,Thomas1,Reilly,Kristensen}. We remind that  
we use the unit of length $a_B \approx 10^{-2}\mu m$, so the typical length 
of barriers presented in Fig.3 and Fig.5 is between $0.1\mu m$ and $0.2\mu m$. 

To understand the reason for the structures in the transmission probability 
we plot the effective selfconsistent  potential $U_{eff}$ 
(see eq. (\ref{Hhf})) and the electron density above the barrier at 
parameters corresponding to these structures.
The point A in Fig.5 indicates the shoulder at $T\approx 0.65$ for the 
barrier of the length $l=10$, the parameters are: $l=10$, $d=1$, $U_0=9.8$.
The corresponding selfconsistent potential is shown in Fig.7 by the
solid line, the external potential (\ref{UU}) is shown by the dashed line.
\begin{figure}[h]
\vspace{-10pt}
\hspace{-35pt}
\epsfxsize=10.cm
\centering\leavevmode\epsfbox{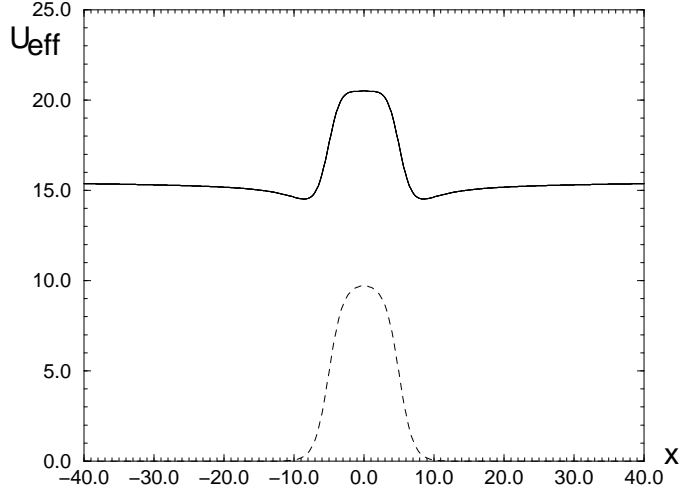}
\vspace{-10pt}
\caption{\it {The selfconsistent HF potential (\ref{Hhf}) is shown
by the solid line.
The external potential (\ref{UU}) is shown by the dashed line. 
The parameters corresponds to the point A in Fig.5:
 $l=10$, $d=1$, $U_0=9.8$.}}
\label{Fig7}
\end{figure}
\noindent
For an estimate the external potential in Fig.7  can be 
approximated near the top by a parabola, $U \sim -\omega_{||}^2 x^2/2$, with
$\omega_{||}\sim 0.5$. This gives the ratio of the typical
longitudinal size to the typical transverse size of the contact, 
$a_{||}/a_t \sim \sqrt{\omega_{\perp}/\omega_{||}} \sim 1$,
see also eq. (\ref{at}).
However the selfconsistent potential, Fig.7,  is very much different from
the simple parabola because of redistribution of electrons along the
contact.
The selfconsistent potential 
calculated in the Hartree approximation is practically the same as 
that calculated in the HF approximation. Nevertheless the HF
transmission probability has a shoulder at the point A, see Fig.5,
and the Hartree transmission probability has no a shoulder, see Fig.6.
The difference is due to the exchange interaction. 
To demonstrate this we plot in Fig.8 the electron density above the barrier.
The HF electron density is shown by the solid line, and
the Hartree electron density is shown by the dashed line.
We see that the shoulder in the transmission probability is related to the
charge density wave above the barrier. There is no a charge density wave in 
the Hartree approximation, and there is no a shoulder in the corresponding 
transmission probability shown in Fig.6.
\begin{figure}[h]
\vspace{-10pt}
\hspace{-35pt}
\epsfxsize=10.cm
\centering\leavevmode\epsfbox{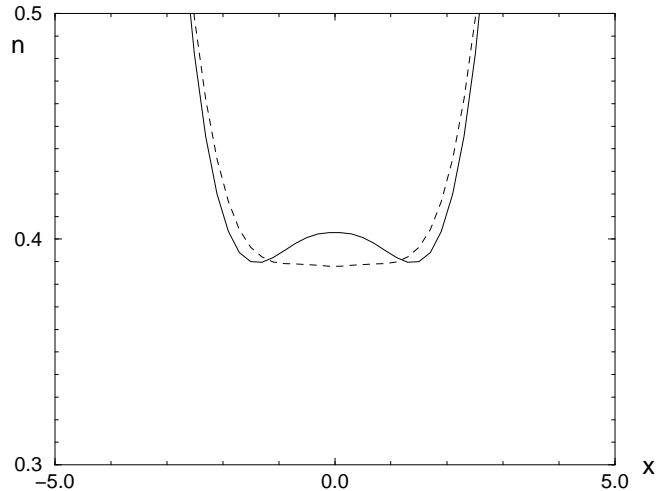}
\vspace{-10pt}
\caption{\it {The electron density on the barrier
at $l=10$, $d=1$, $U_0=9.8$. This corresponds to the
shoulder A in Fig.5. The solid line is obtained in the Hartree-Fock 
approximation and the dashed line is obtained in the Hartree approximation.
}}
\label{Fig8}
\end{figure}
\noindent

The point C in Fig.5 indicates the local minimum at $T\approx 0.5$ for 
the barrier of the length $l=16$, the parameters are: 
$l=16$, $d=1$, $U_0=11.4$.
The corresponding selfconsistent potential is shown in Fig.9 by the
solid line, the external potential (\ref{UU}) is shown by the dashed line.
\begin{figure}[h]
\vspace{-10pt}
\hspace{-35pt}
\epsfxsize=10.cm
\centering\leavevmode\epsfbox{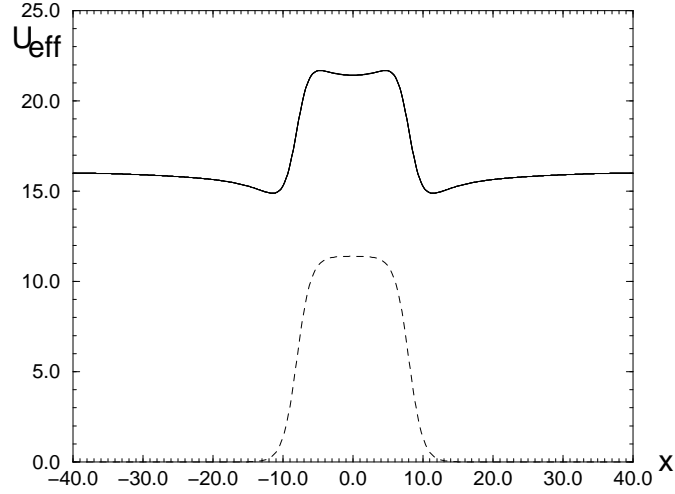}
\vspace{-10pt}
\caption{\it {The selfconsistent HF potential (\ref{Hhf}) is shown
by the solid line.
The external potential (\ref{UU}) is shown by the dashed line.
The parameters corresponds to the point C in Fig.5:
$l=16$, $d=1$, $U_0=11.4$.}}
\label{Fig9}
\end{figure}
\noindent
Similar to the previous case the HF selfconsistent potential
is practically indistinguishable from the Hartree one. 
Nevertheless the HF transmission probability has a well pronounced
minimum, see point C in Fig.5, and
the Hartree transmission probability has just a shoulder, see Fig.6.
The difference is due to the exchange interaction. 
The HF electron density is shown in Fig.10 by the 
solid  line, and the Hartree electron density is shown by the dashed line.
There is an additional modulation of the HF density due to the exchange 
interaction.
\begin{figure}[h]
\vspace{-10pt}
\hspace{-35pt}
\epsfxsize=10.cm
\centering\leavevmode\epsfbox{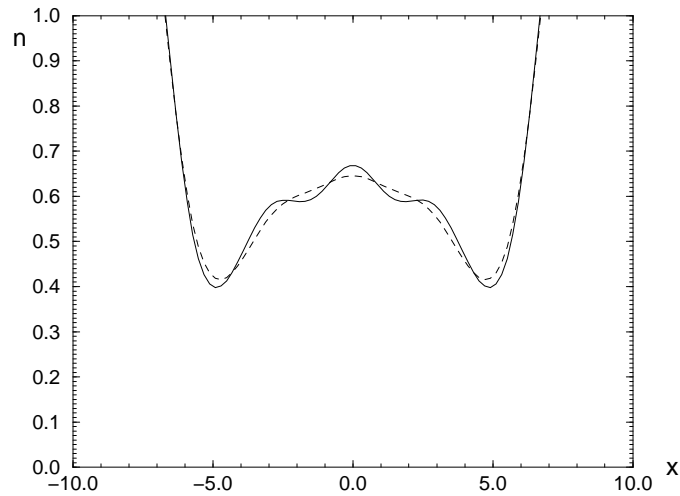}
\vspace{-10pt}
\caption{\it {The electron density on the barrier
at $l=16$, $d=1$, $U_0=11.4$. These parameters correspond to the local
minimum C in Fig.5. The solid line corresponds to the Hartree-Fock 
approximation and the dashed line corresponds to the Hartree approximation.
}}
\label{Fig10}
\end{figure}
\noindent

We see that the structures in the transmission probability are
related to the resonance structures above the potential barrier.
The resonance structures are related to the charge density waves 
developing on the barrier when the average electron density above the 
barrier is of the order of $0.5/a_B$.

\section{Conclusions}
The fictitious gauge field method has been developed to calculate a
potential barrier transmission probability $T$ with account of many-body
effects. The transmission probability is directly related to the
conductance, $G={{2e^2}\over{h}}T$.
The results of the Hartree-Fock calculations of the
transmission probability are shown in Fig.3 and Fig.5.
The results  demonstrate a plateau or a shoulder at the value of the
transmission probability $T \approx 0.7$. For longer barriers 
this structure is getting weak, but additional 
structures appear at lower values of $T$. 
All the structures are related to the development of the
charge density waves on the barrier. This is a precursor for Wigner 
crystallization.
We believe that this explains the conductance anomalies
observed in experiments with quantum contacts 
\cite{Thomas,Thomas1,Reilly,Kristensen}.

We would like to note that there are two further questions that can
be studied within the developed fictitious gauge field approach. The first 
question is how the structures in the conductance evolve in
the external magnetic field. The experimental data \cite{Thomas1} indicates
that the conducatnce anomaly  is somehow related to the spin.
In the picture considered in the present work the effect is certainly
spin dependent because it is due to the exchange interaction.
To analyze this problem in details one should perform calculations 
at $N_{\uparrow}\ne N_{\downarrow}$, so
the problem is more technically involved compared to the considered case. 
Another question is the temperature dependence of the anomalies.
It is known from experiment that the effect is increasing with 
temperature up to $\sim 1K$. Unfortunately there is no a direct way
to consider the finite temperature case within the developed
formalism. However it is clear that any temperature dependence
is related to the excitations above the barrier. Moreover there is
a direct spectroscopic observation of such excitation \cite{Kristensen}.
The problem of the collective excitation can be addresses within the
developed formalism. To approach this problem 
one should apply the time dependent Hartree-Fock method in the external
gauge field. This calculation is substantially
more technically involved compared to the considered case. 

I am grateful to P. E. Lindelof who has attracted my attention to this
problem. I am also grateful to
V. V. Flambaum,  C. J. Hamer, M. I. Katsnelson, M. Yu. Kuchiev, A. Nersesyan,
D. J. Reilly, and J. Thakur for helpful conversations.



\begin{references}
\bibitem{DL} For example if $x=0.1L$ and $y=0.5L$ then $D=y-x=0.4L$,
however if $x=0.1L$ and $y=0.9L$ then $D=0.2L$.
\end{references}
\end{document}